\documentclass[preprint,review,3p,12pt]{elsarticle}

\usepackage{amsmath}
\usepackage{graphics}

\journal{Social Networks}
\usepackage{amssymb}
\usepackage{setspace}
\usepackage{multirow} 
\usepackage[table]{xcolor}
\begin{document}
\begin{frontmatter}
\title{Leadership Network and Team Performance in Interactive Contests}
 \author{Satyam Mukherjee }
\ead{satyam.mukherjee@gmail.com}
\address{Indian Institute of Management Udaipur, Rajasthan, India}

\begin{abstract}
Over the years, the concept of leadership has experienced a paradigm shift $-$ from solitary leader (centralized leadership) to de-centralized leadership or distributed leadership. This paper explores the idea that centralized leadership, as earlier suggested, negatively impacts team performance. I applied the hypothesis to cricket, a sport in which leaders play an important role in team's success. I generated batting partnership network and evaluated the central-most player in the team, applying tools of social network analysis. Analyzing $3420$ matches in one day international cricket and $1979$ Test matches involving $10$ teams, I examined the impact of centralized leadership in outcome of a contest. I observed that the odds for winning a one day international match under centralized leadership is $30$\% higher than the odds for winning under de-centralized leadership. In both forms of cricket (Test and one day international ), I failed to find evidence that distributed leadership is associated with higher team performance. These results suggest important implications for cricket administrators in development and management of working teams. 
\end{abstract}

\begin{keyword}
Social network analysis, Centralized leadership, de-centralized leadership, Sports
\end{keyword}

\end{frontmatter}

\section*{Introduction}
There exists a corpus of work about the benefits of working in teams, a trend which is gaining importance. In academia, it has been shown that works with highest scientific impact have been produced by teams\cite{Guimera,mukherjeesci}. Team coordination is also prized in sports \cite{fewell, kniffin2010, kniffin2005, wolfe2005} and military \cite{hutch, levi}, where team members coordinate with each other for a common objective of being more successful than the opponent. A recent survey conducted on high-level managers concluded that teams are central to organizational success\cite{martin}. The effect of leadership on team performance has been a topic of interest for a long time. Previous works on leadership have dealt with role of leadership in coaching related activities\cite{manz,wageman} or managing events in context of teams\cite{druskat}. Some works have also focussed on how leadership is shared in teams\cite{carson,hiller,pearce}. However, earlier body of work on effect of leadership on team performance was conducted at the level of survey analysis and narrow set of leadership activities\cite{burke,zaccaro}. One of the major drawback of such studies is that team performances were assessed in a subjective manner in which team leaders rated the performance of their own teams. An earlier work has shown that team leaders tend to over-rate team performance, since a team's performance reflects the ability of the leader \cite{sparrow}.

The decisive role of leaders in team's performance has been a long debated topic\cite{smallbusiness, smallbusiness2}. Prior works focussed on the paradigm of leader-centeredness, in which the leadership is viewed as a top-down process between the leader and the followers\cite{yukl}. Recent works have also focussed on the idea of shared leadership or distributed leadership in which other team members emerge as leaders \cite{amehra}. An earlier meta-analysis {of $37$ studies of teams in natural contexts} discus how the network position of team leaders influences team performance\cite{balkundi2006}. {\bf It was observed that teams with stronger interpersonal ties are more successful and teams with leaders who are central in the intra-group networks display better performance \cite{balkundi2006}}. One of the main limitations of the earlier studies is that they are restricted to cross-sectional data, primarily due to the limited availability of longitudinal data. To overcome the limitations of previous works, I employed the treasury of data available in sports \cite{mukherjeecric,Duch,Radicchi, kniffin2014} and objectively investigate the association between leadership structure and  team performance in interactive contests. 

I applied the social network analysis approach to diagnose the role and qualities of a leader effectively. {\bf Leadership involving team activities is a relational construct. Again, social network analysis emphasizes on the relationship of social actors and subsequently elucidates the patterns and theories of such relationships \cite{mayo}. Network analysis has been applied to explore the significance of structure of various relationship in organizations \cite{Krack, Krack2}. Social network approach to leadership demonstrated how would-be leaders perfectly perceives the relationship among team members in various organizations \cite{Balkundi2005}}. Social network analysis provides an understanding of the dynamics of centralized leadership and distributed leadership\cite{amehra,morgeson}. Here, I quantified the extent to which leadership potentials are associated with games won across all teams in the history of cricket. Even though cricket is the second most popular game in the world after soccer, compared to other professional sports it has been relatively understudied by academics, although there is no dearth of match statistics. 

Cricket is chosen for the following reasons. First, cricket is a game in which an outcome depends a lot on the leadership. Compared with other sports the role of a captain is elevated in cricket. {\bf A cricket captain's direct involvement in the proceedings of a game can be viewed as team-leadership in the corporate world, leadership in politics, social capital \cite{Lin99} or organizational communication tactics \cite{Yamac2009}}. The captain chooses the batting order, sets up fielding positions and shoulders the responsibility of on-field decision-making and is also responsible at all times for ensuring that play is conducted within the Spirit of the Game as well as within the Laws. However, a coach in soccer or manager in baseball takes decisions off the field, which includes player substitution or deciding batting line-up. In cricket, the role of a captain is not restricted to off-the-field decisions but also to deliver winning performance for the team while playing \cite{cricketpsych}. It is to be noted that in cricket, there is no substitution unlike Soccer or Basketball, where a player is substituted by the coach. To quote Sir Don Bradman {\it ``A captain must make every decision before he knows what its effect will be, and he must carry the full responsibility, not whether his decision will be right or wrong, but whether it brings success"} \cite{bradman} . 

In cricket, the captains are appointed based on their performance and position in the team (often the role is given to batsmen). One of the key role performed by the captain is leading by example \cite{cricketpsych}, a quality that is gaining importance in business domains \cite{smallbusiness, smallbusiness2}. The captain is expected to win the match for his team, commonly referred by fans and commentators as `captain's knock'. Legendary players like Sir Don Bradman, Richie Benaud or Sir Gary Sobers, were great performers and inspired their team through their own performance $-$ example of centralized leadership. Even though in cricket there are always formally appointed captains, the emergence of leaders has been seen in many games. These emergent leaders were responsible for leading their team to victories. While captains like Mike Brearley or Ray Illingworth were not the best players in their side but were known to extract maximum performance from their players. Again, Sir Gary Sobers and Sachin Tendulkar were best players in their sides, they were not successful captains. In an earlier study it was shown that Steve Waugh was the most successful captain in the history of Test cricket ($1877-2010$)\cite{mukherjeeskip}. Again, presence of legendary performers like Adam Gilchrist, Shane Warne, Glenn McGrath and Ricky Ponting in Steve Waugh's Australian team, leads to the well debated topic whether distributed leadership is more successful than centralized leadership. Secondly, in a team game like cricket, one can objectively assess the role of leader-position in the network and team performance. Motivated by the above observations I set to explore the role of leaders in a team game like cricket and the impact of leadership structure on the outcome of a match.  

\section*{Materials and Methods}

\subsection*{Data}
I analyzed the data of batting partnership (publicly available in cricinfo website \cite{cricinfo}) in Test cricket between $1877$ and $2013$ and also one day international cricket between $1971$ and $2013$. Cricinfo has recorded the information for all $3420$ one day international  matches played between $1971$ and $2013$ and all $1979$ Test matches played between $1877$ and $2013$. For every match I recorded and analyzed the score-cards which contain the information of match outcome, amount of runs scored by a pair of batsmen and run-rate of each team after the game is over. %Run rate is defined as the amount of runs scored by a team divided by the number of overs consumed in scoring those runs. 
In order to control for team talent, I also collected the information about the International cricket Council (ICC) points awarded to every player each year as well as the batting average of every player (including the captain) in a year. %The ICC points serve as an indicator of the talent of the players. 
Data are available upon request \footnote{The author will share the data in an online repository post publication of the manuscript}. 

\subsection*{Network Representation}
\begin{figure}[!ht]
\begin{center}
\includegraphics[width=6in]{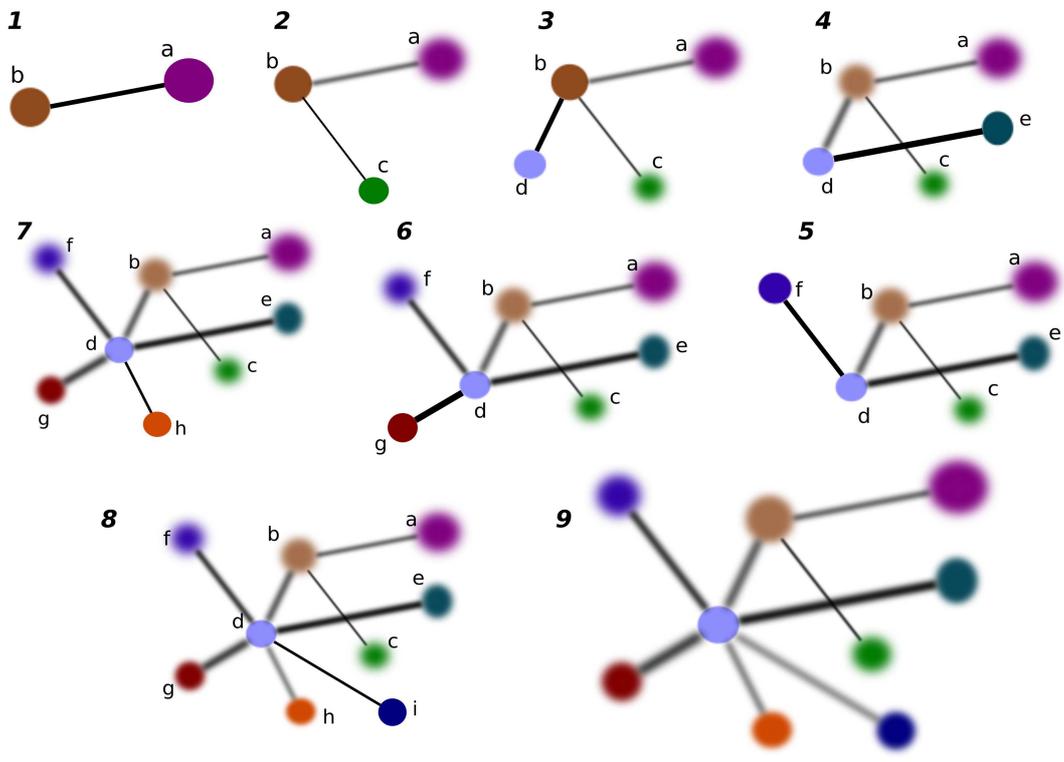}
\end{center}
\caption{{\bf Batting partnership network} The batting pair who start the innings is referred to as an opening-pair. Two opening batsmen $a$ and $b$ start the innings. A weighted link is generated between $a$ and $b$. The weight of the link is proportional to the runs scored by the batting pair ($a$, $b$) ({\bf 1}). Now, if batsman $a$ is dismissed by a bowler, then the partnership between $a$ and $b$ is broken and a new batsman $c$ arrives to form a new partnership with batsman $b$ ({\bf 2}).
If $c$ is dismissed then a new batsman $d$ gets linked with batsman $b$ ({\bf 3}). In this way one can generate an entire network of batting-partnership till the end of an innings (${\bf 4}-{\bf 9}$). The nodes and links are blurred if a batsman is dismissed and a partnership is broken.}
\label{fig:fig0}
\end{figure}

To articulate the social network analysis approach of studying the pattern of leadership in cricket, I first outline the methodology of identifying the leadership style between two competing teams. Next I discuss the nature of leadership networks and finally discuss the effect of centralized and distributed leadership on the outcome of a game. In cricket two batsmen always bat in partnership, although only one is on strike at any time. The partnership of two batsmen comes to an end when one of them is dismissed or at the end of an innings. Fig~\ref{fig:fig0} demonstrates the formation of batting partnership network. 

Two opening batsmen $a$ and $b$ start the innings for their team. In network terminology, this can be visualized as a network with two nodes $a$ and $b$, the link representing the partnership between the two players. Weight of the link reflects the amount of runs scored in partnership. Now, if batsman $a$ is dismissed by a bowler, then a new batsman $c$ arrives to form a new partnership with batsman $b$. 
Thus a new node $c$ gets linked with node $b$. Subsequently one can generate an entire network of batting-partnership till the end of an innings. The innings comes to an end when $10$ players are dismissed or when the duration of play comes to an end\cite{mukherjeebpn}. The score of a team is the sum of all the runs scored during a batting partnership. 

The outcome of a match depends on the batting partnerships between batsmen. Long lasting partnerships not only add runs on the team\rq{}s score, it may also serve to exhaust the tactics of the fielding team. Again, the concept of partnerships become vital if only one recognized batsman remains. It is therefore important to identify the key players in a team by constructing network of batting partners. Two batsmen are connected if they formed a batting partnership in the match. An undirected and weighted batting partnership network is generated for each team and for every match played through $2013$. I examined two network metrics which captures the position of a captain in the batting partnership network in cricket. A similar approach using network metrics to capture team performance and strategies has been used in basketball \cite{fewell} and soccer \cite{tgrund}. 

\paragraph{ Centralized leadership and de-centralized leadership} To quantify the centrality of a captain, I evaluated the betweenness centrality of players in the batting partnership network. The betweenness centrality is defined as 
\begin{equation}
C^{w}_{B}(i) = \frac{g^{w}_{jl}(i)}{g^{w}_{jl}}
\end{equation}
Where $w$ is the weight of the link between two nodes $j$ and $l$, $g_{jl}$ is the number of shortest paths between two nodes and $g_{jl}(i)$ is the number of shortest paths that pass through node $i$ \cite{wasserman, peay}.  
Betweenness centrality measures the extent to which one batsman is between two other batsmen who are not connected to each other. In other words, betweenness centrality measures how the run scoring by  a player during a batting partnership depends on another player. Batsmen with high betweenness centrality are crucial for the team for scoring runs without losing his wicket. These batsmen are important because their dismissal has a huge impact on the structure of the network\cite{pena}.  So a single player with a high betweenness centrality is also a weakness, since the entire team is vulnerable to the loss of his wicket. In an ideal case, every captain would seek a combination of players where betweenness scores are uniformly distributed among players. Hence betweenness centrality is a measure of dependence on other team members\cite{pena}. Centralized leadership refers to the post-match situation when captain is the player with highest betweenness centrality, else it is an example of emergent de-centralized leadership. 

\paragraph{ Distributed leadership} To predict a continuos measure of leadership structure I measure the network de-centralization proposed by Mayo et al \cite{mayo}. The variance of centrality is given by the equation
\begin{equation}
\omega =  \frac{\sum_{i=1}^N (k_{max} - k(i))}{(N-1)(N-2)}
\end{equation}
Where $N$ is the number of players in the batting partnership network and $\omega$ is the variance of centrality of the network and $k$ is the degree of the node, with $k_{max}$ being the maximum degree. Here degree of a node refers to the number of batting partners of a player. Variance of centrality ($\omega$) captures the aspects of batting performance of players under the leadership of different captains. As mentioned earlier the captain takes the bulk of the decision in forming the batting line-up. A good leader will always allot effective batting positions such that the team is benefited the most. In this sense the $\omega$ highlights the extent to which the leadership is distributed. The index $\omega$ varies from $0$ to $1$. A value of $0$ indicates that the leadership is distributed equally among the individuals and a value of $1$ indicates that the team is centralized around an individual, not necessarily the captain.

\paragraph{Normal approximation method of the Binomial confidence interval} The equation for Normal approximation method\cite{seana} to evaluate 95\% Binomial confidence intervals is given as,
\begin{equation}
CI =  p~ \pm~1.96~\sqrt\frac{p(1-p)}{M}
\end{equation}
Where $p$ is the proportion of interest and $M$ is the number of matches played. For example if out of $M$ matches, $m$ matches results in win under centralized leadership, then the 95\% Binomial confidence intervals fall between $\frac{m}{M} - 1.96~\sqrt\frac{\frac{m}{M}(1-\frac{m}{M})}{M}$ and $\frac{m}{M} + 1.96~\sqrt\frac{\frac{m}{M}(1-\frac{m}{M})}{M}$. 

\subsection*{ Regression Methods and Control variables}
Below I report detailed results for predicting the probability of win using logistic regression analysis.  The dependent variable is a binary indicator for win$-$loss. The explanatory variable of interest is an indicator variable of centrality of the captain, as defined above. The control variables are as follows:

\paragraph{ Batting average of captain} Batting average of a player is defined as the amount of runs scored divided by the number of times the player is dismissed. Batting average of a captain serves as an indicator of his ability and skills as a batsman irrespective of external factors like match situation or strength of opposition. In the logistic regression analysis, I include a dummy variable which takes the value $1$ if the captain's batting average is higher than the median batting average of the team; $0$ otherwise. 

\paragraph{ Talent of captain} While the batting average of a player serves as a good metric of a batsman's ability, it is independent of match situations, quality of bowling attacks, or strength of opponents. The ICC player rankings provides a sophisticated ranking of batsmen based on amount of runs scored, quality of opposition, winning performance for the team. Players are rated on a point scale of $0$ to $1000$, more points being granted if the opponent is stronger or the player's performance results in team's win. The ICC points of the captain is an indicator of his batting talent. In the logistic regression analysis, I introduce a dummy variable which takes the value $1$ if the captain's ICC points is higher than the median ICC points of the team; $0$ otherwise. 

\paragraph{ Fixed effects} The logistic regression includes a full set of team fixed-effects, fixed effects of each year the match was played and fixed effects of batting position of the captain during the match.

The logistic regression takes the form,
\begin{equation} 
logit(W_{i}) = \beta_{1}~C(i) + \beta_{2}~S_{b}(i) + \beta_{3}~S_{p}(i) + {\sum_{t=1}^m \gamma_{t}}~Team_{ti} + {\sum_{y} \gamma_{y}}~Year_{yi}  + {\sum_{p} \gamma_{p}}~Pos_{pi}\, 
\end{equation}
Where $C(i)$ is an indicator variable for centrality of captain, $W(i)$ is an indicator variable of win-loss by a captain. The indicator variable $S_{b}$ ($S_{p}$) takes a value $1$ if the batting average (ICC points) of the captain is above the median batting average (median ICC points) of the team; $0$ otherwise. The logistic regression includes a full set of fixed effects for each of teams, where the indicator variables $[(Team)]_{ti}$ $\in$ \{0,1\}  are equal to $1$ if the match $i$ involves team $t$. I also controlled for fixed effect of the year the match was played - indicator variables $[(Year)]_{yi}$ $\in$ \{0,1\}  are equal to $1$ if the match $i$ was played in the year $y$ and {\bf fixed effect of the batting position of the captain - $[(Pos)]_{pi}$ $\in$ \{0,1\}  are equal to $1$ if the batting position is $p$ in match $i$}. 

\subsubsection*{Distributed leadership model} The dependent variable is the difference of run-rates of a team, defined as the ratio of number of runs scored by a team to the total number of overs played by the team $-$ if a team scored $140$ runs in $20$ overs, run-rate is $7$. The explanatory variable is the difference of variance of centrality of two teams. The control variables include difference in batting average of captains as well as difference in team talent of competing teams. Team talent is measured as the coefficient of variation of ICC points of every player in the team. The linear model takes the form,
\begin{equation} 
\delta~r_{12}(i) = A_0 + A_1~\delta~\omega_{12}(i)  + A_2~\delta~C^{v}_{12}(i) + A_3~\delta~B^{Avg}_{12}(i) + A_4~{\sum_{g=1}^m \gamma_{g}}~Ground_{gi} + A_5~{\sum_{y} \gamma_{y}}~Year_{yi} \, 
\end{equation}
Where the dependent variable $\delta~r_{12}$ is the difference of run-rate of the two competing teams ($r_{1} - r_{2}$), $\delta~\omega_{12}$ is the difference of degree centralities of the two teams ($\omega_{1} - \omega_{2}$), $\delta~C^{v}_{12}$ is the difference in coefficient of variation of team talent (ICC points) of the teams and $\delta~B^{Avg}_{12}$ is the difference in batting average of the captains leading the two teams. The subscripts $1$ and $2$ indicates the first and second batting innings. The regression includes a full set of fixed effects for each games, where the indicator variables $[(Ground)]_{gi}$ $\in$ ${(0,1)}$  are equal to $1$ if the game $i$ was played in ground $g$. 

\section*{Results}
\paragraph{ Centralized leadership and team performance} I empirically investigated whether team performance is negatively associated with centralized leadership. If the captain emerge as the player with highest betweenness centrality, then the team is under centralized leadership. On the other hand, in many cases, other players may emerge as most central player $-$ an example of emerging de-centralized leadership. In Fig~\ref{fig:fig1} I compared the batting partnership network of two competing teams $-$ {\it England} and {\it Australia} during the first Test match in Ashes $2013/2014$. The Australian captain {\it M.~J.~Clarke} is not the central-most player, whereas  for England, captain {\it A.~N.~Cook} emerged as the player with highest betweenness centrality. The first Test match resulted in a defeat for England $-$ an example where centralized leadership is negatively associated with team performance. I analyzed whether {\bf the first Test match in Ashes $2013/2014$ is a special case and whether} leadership structure has any relationship with team performance in cricket matches. For each game the betweenness centrality of captains in the batting partnership network is evaluated. An indicator variable $C_{i}$ is introduced for centrality of captain $i$, which takes a value $1$ if the captain emerges as the player with highest betweenness centrality and $0$ otherwise.

In Test cricket, there are three possible outcomes $-$ win, loss or draw. The success or failure of a team is directly related to the success or failure of the team's captain. In order to assess a team's success, I assigned a score of $0$ if the team loses, a score of $1$ if the match is drawn (indecisive) and score of $2$ if team wins the match. In the entire history of Test cricket there has been only two games which resulted in tie. Here I assigned the score of $1$ for tied matches. In one day international  cricket there are three possible outcomes $-$ win, loss or tie. As mentioned earlier a score of $0$ was assigned if the team loses, a score of $1$ if the match was tied and score of $2$ if the team won the match. The average of all the scores for each team is an indicator of the performance of the captain. Fig~\ref{fig:fig2} shows that in Test cricket, the leadership structure doesn't have a significant effect on team performance. Interestingly, in one day international  cricket a significant advantage is seen for centralized leadership ($p < 0.001$). 

Team performance can be assessed if the captain is the most central player ($C(i)=1$) or captain is not the most central player ($C(i)=0$) at the end of a game. I hypothesize that the centrality of the captain has a stronger effect on the outcome of a game compared to the batting average and talent of the captain. To assess the robustness of the association between leadership structure and the team performance, the relationship is quantified with a logistic regression of the form $logit(W_{i}) = \beta_{1}~C(i) + \beta_{2}~S_{b}(i) + \beta_{3}~S_{p}(i) + {\sum_{t=1}^m \gamma_{t}}~Team_{ti} + {\sum_{y} \gamma_{y}}~Year_{yi}$ (Materials and Methods). 
As summarized in Table~\ref{table_regression_1}, I observe that in one day international  cricket, the probability of winning depends positively and significantly on centrality $C$ of the captain ($p<1\times 10^{-16}$), negatively depends on the batting average as well as captain's talent. For the latter variable, no significant relationship is observed with the outcome of a match. The odds of winning a one day international match for centralized leadership ($C=1$) over the odds of winning for de-centralized leadership ($C=0$) is $\exp~(0.262)$ $=1.299$. In other words, the odds for centralized leadership is $30$\% higher than the odds for de-centralized leadership. The same logistic analysis reveals that in Test cricket, captain's centrality, captain's batting average, and captain's talent are statistically unrelated to the performance of the team ($p > 0.05$). 

\paragraph{ Binomial confidence interval} In an alternate approach, I implemented a non-parametric test to evaluate the confidence of the results. I estimated the number of wins when the captain is the player with highest centrality as well as number of wins when the captain is not the player with highest centrality and evaluate the Binomial Confidence Interval (BCI) using the Normal Approximation Method\cite{seana}. In one day international  cricket, out of $891$ matches, $459$ resulted in win when the captain is the player with highest centrality, with the $95\%$ BCI falling between $48.2\%$ and $54.8\%$. Considering the one day international  matches where the captain is not the player with highest centrality, I observe that the $95\%$ BCI falls between $43.8\%$ and $46.6\%$ ($2293$ wins out of $5067$ matches), strengthening the hypothesis that centralized leadership is more successful than de-centralized leadership. However, in Test cricket I don't observe any significant difference between success under centralized leadership or de-centralized leadership - the $95\%$ BCI falling between $61.5\%$ and $69.8\%$ for centralized leadership ($338$ wins and draws out of $514$ matches) and between $65.6\%$ and $68.7\%$ for de-centralized leadership ($2360$ wins and draws out of $3512$ matches).

\paragraph{ Distributed leadership and team performance} So far I described centralized leadership and de-centralized leadership based on the betweenness centrality of the captain in batting partnership network.  In emergent de-centralized leadership, a player other than the formally appointed leader emerge as the most central player and takes the responsibility for winning performance for the team. {\bf  In such situations, team members other than the captain also score high as well}. There also exists situations where different team members rotate to take leadership responsibilities - an example of distributed leadership, a team level construction \cite{derue2011}. As defined in Equation $2$, I utilized the variance of centrality as an indicator for distributed leadership. Next, I investigated the impact of leadership structure on team performance. The team performance is judged by the run-rate of the team. A higher run-rate indicates a superior performance of the team. In a team game, performance of the team depends on the team talent \cite{swaabr}. As considered earlier, the ICC points for every player in a team serves as a quantifier for overall team talent (Materials and Methods). I collected the ICC points for every player in a team per year and evaluate the coefficient of variation of the ICC points. If I hypothesize that distributed leadership positively affect the outcome of a game, then it would follow that the difference of variance of centrality of two teams is negatively associated with the difference of run-rate of teams. Controlling for the team talent and batting average of the captain, the relationship between difference in team run-rate and difference in variance of centrality is quantified by a linear model mentioned in Equation $5$ (Materials and Methods).

The linear model indicates that in one day international  cricket team talent ($\delta~C^{v}_{12}$) and captain batting average ($B^{Avg}_{12}$) have no significant explanatory power for $\delta~r_{12}$. Consistent with the conjecture, there exists a positive and significant relationship between $\delta~r_{12}$ and variance of centrality ($\delta~\omega_{12}$) in one day international  cricket (Table~\ref{table_regression_2}). Together, they account for about $10\%$ of the observed variance in the data. No significant association was observed between $\delta~r_{12}$ and $\delta~\omega_{12}$ in Test cricket ($p=0.883$), although team talent plays a significant effect on the result ($p<1\times 10^{-16}$). In Test cricket, variance of centrality, team talent and captain batting average account for only $4\%$ of the observed variance in the data.

Figure~\ref{fig:fig3} shows the standardized coefficients for one day international and Test cricket and compares the relative effects of the variables $-$ $\delta~\omega_{12}$,  $\delta~B^{Avg}_{12}$ and $\delta~C^{v}_{12}$. In one day international matches the relative effect of the difference of variance of centrality is significantly higher than the difference of batting average of captain and difference of coefficient of variation of ICC points of team $-$ a one standard deviation increase in $\delta~\omega_{12}$ yields a $0.269$ standard deviation increase in the predicted difference of team run-rate. In Test cricket, difference in coefficient of variation of ICC points (team talent) has significantly higher effect than leadership structure or difference in batting average of captains of two teams $-$ one standard deviation decrease in $\delta~C^{v}_{12}$ results in $0.097$ standard deviation increase in $\delta~r_{12}$.

\section*{Discussion}
\begin{figure}[!ht]
\begin{center}
\includegraphics[width=6in]{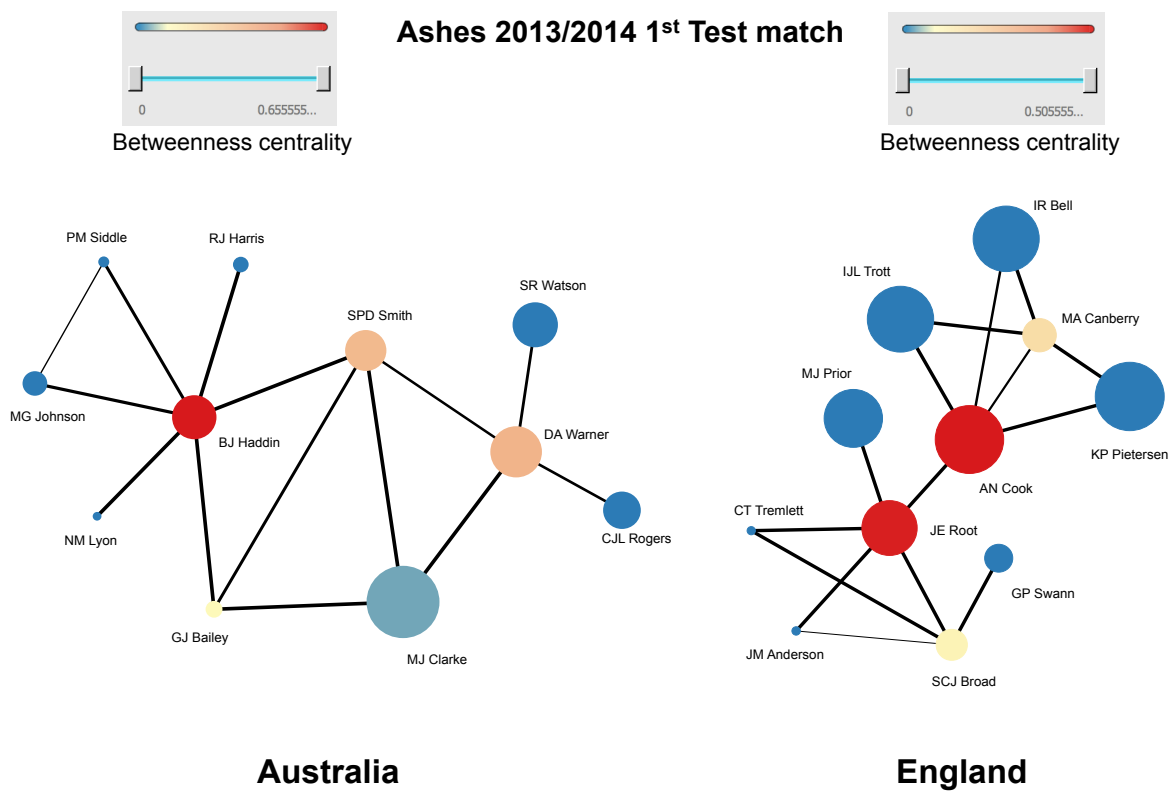}
\end{center}
\caption{{\bf Batting partnership network for Australia and England} Batting partnership network of batsmen playing under the leadership of {\it M. J. Clarke} (Australia) and {\it A. N. Cook} (England). The nodes represent the batsmen and the links denote the batting partnership. Each node is colored according to the betweenness centrality of the node. Thickness of each link if proportional to the runs scored by a batting pair for a given team. The size of each node is proportional to the career batting average of the player. }
\label{fig:fig1}
\end{figure}

\begin{figure}[!ht]
\begin{center}
\includegraphics[width=6in]{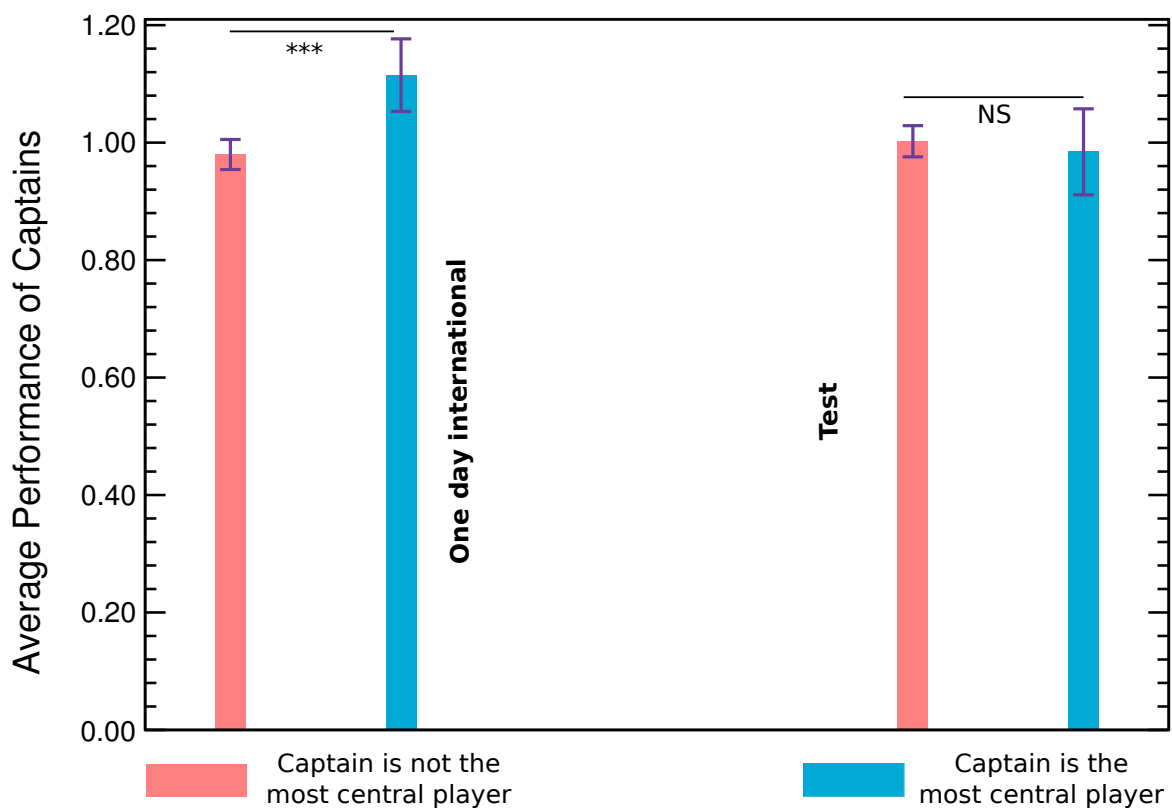}
\end{center}
\caption{{\bf Average performance of teams under centralized and de-centralized leadership} A team is assigned a score of $0$ if the team loses, a score of $1$ if the match is drawn (indecisive) and score of $2$ if team wins the match. The average of all the scores for each team is an indicator of the average performance of the captain. I plot the average points of teams when the captain is the most central player of the team (blue) and when the captain is not the most central player of the team (red). In one day international  cricket matches centralized leadership shows significant advantage over distributed leadership. However in Test cricket, no significant difference is observed between centralized leadership and de-centralized leadership. Here, bars represent $95 \%$ confidence intervals obtained by resampling, $^{***}$ denotes $p < 0.001$ and $^{NS}$ denotes not significant. }
\label{fig:fig2}
\end{figure}

\begin{figure}[!ht]
\begin{center}
\includegraphics[width=6in]{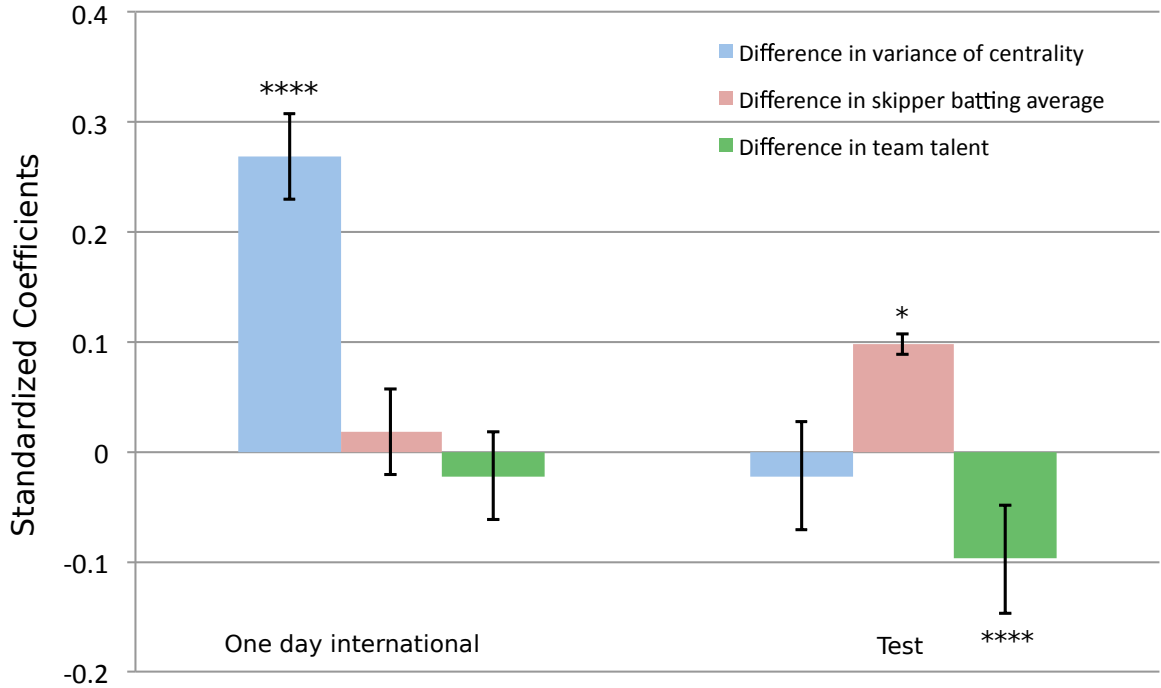}
\end{center}
\caption{{\bf Relative effect of $\delta~\omega_{12}$, $\delta~C^{v}_{12}$ and $\delta~B^{Avg}_{12}$ terms on difference of team run rates} Figure shows the standardized coefficients with 95\% confidence intervals and compare the relative effects of the variables $\delta~\omega_{12}$, $\delta~C^{v}_{12}$ and $\delta~B^{Avg}_{12}$. Significant variables are marked by (*), where  * $p < 0.05$; **** $p < 1\times 10^{-16}$. In one day international cricket, a one standard deviation increase in $\delta~\omega_{12}$ yields a $0.269$ standard deviation increase in the predicted difference of team run-rate. Note that there is no significant relationship between the control variables $\delta~C^{v}_{12}$ and $\delta~B^{Avg}_{12}$ with the difference in team run-rates. In Test cricket, there is no significant relationship between the explanatory variable $\delta~\omega_{12}$ and the dependent variable $\delta~r_{12}$. However, a one standard deviation increase in $\delta~B^{Avg}_{12}$ results in $0.098$ standard deviation increase in $\delta~r_{12}$. There also exists significant association between team talent and the dependent variable $\delta~r_{12}$ $-$ one standard deviation decrease in $\delta~C^{v}_{12}$ results in $0.097$ standard deviation increase in $\delta~r_{12}$.}
\label{fig:fig3}
\end{figure}

Contribution of this paper is of practical importance in research involving leadership perception in teams. While effect of leadership on team performance has long been analyzed under the premises of survey analysis, an extensive empirical evidence was lacking. {\bf Contrary to the example discussed in Fig~\ref{fig:fig1}}, at least in the context of Test cricket, my results extensively demonstrates that there is no evidence to suggest distributed leadership as well as de-centralized leadership is associated with better team performance in competitions involving small teams in cricket matches. The magnitude of the coefficients enable us to infer that, in one day international cricket, centralized leadership shows positive effect on team performance, which can be justified by the fact that one day international cricket is more competitive and result oriented than Test cricket, where there are situations where teams attempt to draw a match and the outcome remains indecisive. These findings perhaps indicate that depending on the level of competitiveness, centralized leadership is positively related with team performance. While this positive correlation cannot establish a causal relationship, it nonetheless suggests strongly the positive relation between centralized leadership and team performance. Contrary to my findings, an earlier study on $28$ field-based sales teams which showed that certain types of distributed leadership (distributed-fragmented leadership) are positively related to team performance \cite{amehra}. {\bf My results confirm the earlier findings of link between centralized leadership and greater team performance, as observed in the meta-analysis of study of $37$ teams in natural contexts \cite{balkundi2006}}. Beyond cricket this approach could be extended to serve as template for analyzing other small team collaborations. It would be interesting to conduct similar research on other professional domains like basketball and soccer in which the most central player is identified by the ball passing networks among players. 

One of the potential limitation of the current work involves the process of captain selection which is an endogenous process. A captain is assigned by a selection committee to maximize the chances of winning. Currently, I am unable to deal with this crucial endogeneity due to lack of available information about the selection process. Nevertheless, these findings leave a lot of potential for future research. For example, one of the key aspects of leadership is experience. It has been shown in earlier works that on average basketball teams with coaches early in their careers benefit relatively more from timeouts than teams with high-experienced coaches\cite{saavmukbag}. Previous research has also shown that in mathematics, mentors early in their careers can have a stronger positive impact on prot\'{e}g\'{e}s than later in their careers \cite{Malmgren}. It remains to be seen if there exists a link between experience and leadership in sports teams, academia or business. While the effect of centralized or distributed leadership in academia or business cannot be demonstrated, preliminary results provide insights into the structure of leadership and performance of cricket teams. It remains an open challenge to extrapolate these findings to domains of scientific and managerial impact. 

Finally there are few more aspects which deserve closer scrutiny. The relationship between gender diversity and decision making hasn't been explored effectively. While an earlier survey of executives observed that there exists a direct connection between gender diversity and business success\cite{guardian}, the connections between success, leadership structure and gender diversity still remains an unexplored challenge. Performance in academia or sports and leadership abilities depending on gender remains an open idea for further research. Again, teams are prone to conflict between team members. A recent study has empirically attempted to predict future conflict in team-members\cite{nuria}. Leadership will play a critical role in diagnosing and resolving potential conflicts and confrontations among team members. 

\paragraph{Limitations} This paper is limited to the situation where captains led by example, specifically identifying whether the most central player in the batting partnership network is captain of the team. Batting partnership network provides a visual summary of how the batting is centralized around an individual batsman or distributed around multiple batsmen. Since the captain is responsible in selecting an effective batting order, batting partnership network reflects a limited set of roles performed by the captain. There are various other roles which are performed by the captain $-$ inspiring and motivating the players they lead, communicating effectively with the coach and selectors, remaining positive in match situations and devising strategies accordingly. {\bf The work is also limited by the fact that I have considered static network of batting partnerships, when the innings come to an end. This limitation is due to the available data in Cricinfo. A more robust analysis is possible by considering the dynamic version of the batting partnership network similar to the idea of ``dynamic exchange networks" \cite{Leik92}, in which the players change as players exchange their batting-striking position. Traditionally, $90$\% of the captains have been specialist batsmen. The present study deals with captains who are part of batting partnership network. However, captains like Courtney Walsh of {\it West Indies}, who are not specialist batsmen, would not end up being the most central player solely due to his low batting order and poor batting abilities. Future analysis involves new techniques to analyze the effect of captains who are specialist bowlers}.  Also, the work is limited to the structural approach to leadership and doesn't explore concepts  like work environment under leadership structure. Whether members in centralized-leadership network experience higher levels of conflict than members of distributed-leadership network is a matter worthy of future investigations.

%\section*{Acknowledgements}
  %The author thanks the cricinfo website for public availability of cricket statistics. 

\clearpage

\clearpage

\begin{table}
\caption{{\bf Results for the logit regression used for predicting the effect of centrality of captain on outcome of a match ($W$)}. Coefficients that are statistically significant (p-value $< 0.05$) are marked in bold fonts. *** $p<0.001$, ** $p<0.01$}
\centering
\begin{tabular}{lc|ccc}

&&& \bf{Model}\\
&&&&\\
&  & &  $\beta$ (SE) &

\\ \hline

        \bf{Model for $W$ in one day } &&\\
        \bf{international  matches} &&\\

&&&\\

	~~~Centrality of captain  &$C$ & \bf{0.251}**(0.072) &  \bf{0.262}**(0.085) &   \\ 	
	&&&\\
	~~~Batting average of captain  &$S_{b}$ & & -0.039 (0.068) &    \\  
	&&&\\
	~~~Batting talent of captain  &$S_{p}$ &  &  \bf{0.006}***(0.0004) &    \\ 

&&&\\
	
{\bf Fixed effects} &&& \\
	~~~Team & & & $Y$ &   \\ 	
	~~~Year  &  & & $Y$ &    \\ 
	~~~Batting position &  &  & $Y$ &    \\ 

&&&\\

        ~~~Prob  $>$ Chi-square& & $0.0006$&$<1\times 10^{-16}$ \\ 
	~~~Number of observations & $5936$ &  &  \\ 
&&&\\      
 \hline

        \bf{Model for $W$ in } &&\\
 	 \bf{Test matches} &&\\
       
&&&\\

	~~~Centrality of captain  &$C$ & $-0.064 (0.099)$ & $-0.014 (0.114)$ &   \\ 	
	&&&\\
	~~~Batting average of captain  &$S_{b}$ & & $0.719 (0.594)$ &    \\  
	&&&\\
	~~~Batting talent of captain  &$S_{p}$ &  &  \bf{0.005}***(0.0006) &    \\ 

&&&\\
	
{\bf Fixed effects} &&& \\
	~~~Team & & & $Y$ &   \\ 	
	~~~Year  &  & & $Y$ &    \\ 
	~~~Batting position &  &  & $Y$ &    \\ 

&&&\\

        ~~~Prob  $>$ Chi-square& & $0.517$&$<1\times 10^{-16}$\\ 
	~~~Number of observations & $4026$ &  &  \\ 

&&&\\      
 \hline
 
&&&&\\ 
\end{tabular}
\label{table_regression_1}
\end{table}

\clearpage

\begin{table}
\small
\centering
\caption{{\bf Results for the linear regression used for predicting the effect of  difference in variance of centrality on difference of run-rate of competing teams in a match.} Bold font is used to mark the coefficients that are statistically significant (p-value$<0.05$).  The variance inflation factor is less than $2$, indicating that severity of multicollinearity affecting the regression is within the tolerance limit. *** $p<0.001$, ** $p<0.01$}
\begin{tabular}{lc|ccc}

\\

&&& \bf{Model}\\
&&&&\\
&  & &  $\beta$ (SE) &

\\ \hline

        \bf{Model for $\delta~r_{12}$ in } &&\\
         \bf{one day international  matches} &&\\

&&&\\

	~~~Difference in variance of centrality &$\omega_{12}$ &   \bf{ 1.066}*** (0.073) & \bf{ 1.009}*** (0.084) &  \\ 		
	~~~Difference in skipper batting average &$B^{Avg}_{12}$ &  & 0.0004 (0.0006) &  \\ 
	~~~Difference in team talent &$C^{v}_{12}$ & &  -0.243 (0.231) &   \\ 
&&&\\
	
{\bf Fixed effects} &&& \\
	~~~Team & & & $Y$ &   \\ 	
	~~~Year  &  & & $Y$ &    \\ 
	~~~Ground &  &  & $Y$ &    \\ 
&&&\\
        ~~~R$^{2}$ &&~ {\bf 0.08} & ~ {\bf 0.106}\\ 
~~~Number of matches & $3420$ &  &  \\ 
&&&\\      
 \hline

        \bf{Model for  $\delta~r_{12}$ in Test matches} &&\\
       
&&&\\

	~~~Difference in variance centrality &$\omega_{12}$ & -0.046 (0.097) & -.015 (0.099) &  \\ 	
	~~~Difference in skipper batting average &$B^{Avg}_{12}$ &  & \bf{0.002}**(0.001) &  \\ 
	~~~Difference in team talent &$C^{v}_{12}$ & &  \bf{ -0.530}**(0.137) & \\ 
&&&\\
	
{\bf Fixed effects} &&& \\
	~~~Team & & & $Y$ &   \\ 	
	~~~Year  &  & & $Y$ &    \\ 
	~~~Ground &  &  & $Y$ &    \\ 
&&&\\

        ~~~R$^{2}$ && $0.0002$& ~ {\bf 0.048} \\ 
~~~Number of matches & $1979$ &  &  \\ 
&&&&\\      
 \hline

&&&&\\      
\end{tabular}
\label{table_regression_2}
\end{table}

\clearpage

\end{document}